\documentclass[prd,aps,showpacs,secnumarabic]{revtex4}
\usepackage{graphicx}
\begin{document}
\def\ds{\displaystyle}
\def\ss{\scriptstyle}
\def\sb{\mbox{\rule{0pt}{8pt}}}
\title{Discriminating between models for the dark energy }
\author{Duane A. Dicus} \affiliation{Center for Particle Physics, University of Texas, Austin,
TX 78712}\author{Wayne W. Repko} \affiliation{Department of Physics and
Astronomy, Michigan State University, East Lansing, MI 48824}

\date{\today}

\begin{abstract}
   Recent measurements suggest our universe has a substantial dark energy component, which is usually interpreted in terms of a cosmological constant.  Here we examine how much the form of this dark energy can be modified while still retaining an acceptable fit to the high redshift supernova data.  We first consider changes in the dark energy equation of state and then explore a model in which the dark energy is interpreted as a fluid with a bulk viscosity.

\end{abstract}
\maketitle

\section{Introduction}

Publication of the analysis of type Ia supernovae red shift data by the High-z
Supernova Search Team \cite{hiz} and the Supernova Cosmology Project
\cite{super} provided the first indication that the universe is accelerating.
The data favor an interpretation in which most of the energy in the universe is some form of dark energy capable of providing negative pressure. The usual
candidate for the dark energy is a cosmological constant $\Lambda$, which, as
the name suggests, gives a time independent dark energy density. Both groups
analyze their data assuming contributions from ordinary matter and a
cosmological constant. They present their results as $\chi^2$ contours the the
plane of the parameters $\Omega_M$ and $\Omega_{\Lambda}$, defined below,
which are a measure of the relative contributions of matter and dark energy
\cite{hiz,super}. From these plots (e.g., Fig.\,\ref{p_w=-1} and Table\,\ref{fits}), it is clear that values of $\Omega_M$ and $\Omega_{\Lambda}$ are most probably those of an accelerating cosmology.

While this picture is attractive because of its simplicity, one can ask to
what extent the data can distinguish between alternative hypotheses for the
behavior of the dark energy. Investigations into this question are generally
framed in terms of determining the equation of state of the dark component
($X$), that is, determining $w=p_X/\rho_X$, where $p_X$ is the pressure of the
dark component and $\rho_X$ is its energy density, as a function of the scale
parameter $R(t)$ \cite{garna,dhmt}. It is possible to quantify the effect of
any particular $w$ by fitting the supernova data and comparing the resulting
value of $\chi^2$ with the one for $w=-1$, which corresponds to the case of a
cosmological constant.

In the next Section, we review briefly the strategy for extracting the
cosmological parameters $\Omega_M$ and $\Omega_{\Lambda}$ from the effective
magnitude data for type Ia supernovae, and use a straightforward $\chi^2$
minimization to reproduce the results of Refs.\,\cite{hiz} and \cite{super}.
In Section \ref{quin}, we use the values of $\chi^2$ for $w=-1$ to explore
several modifications to the equation of state including a search for the minimum $\chi^2$ for the case $w={\rm constant}$. Changes in the pressure-density relation of this type or similar modifications, including models using a Chaplygin equation of state \cite{Fabris:2001tm,Bilic:2001cg,Bento:2002ps,Fabris:2002xx,Bilic:2002vm,
Fabris:2002vu,Dev:2002qa,Gorini:2002kf,Makler:2002jv,Bento:2002uh,Bento:2002yx,
Alcaniz:2002yt,CF} or generalized Cardassian expansion \cite{Freese:2002sq,
Sen:2002ss,Zhu:2002yg}, have been considered before. Our discussion is included here as a prologue to Section \ref{visc}, where we present a model in which the dark component is a fluid with a bulk viscosity. It is shown that this model provides an equally good fit to the supernova data, predicts an accelerating universe and has mass density fluctuations which grow. All of this occurs at the expense of some entropy production. Finally, we conclude with some comments on the current supernova data's capacity to discriminate between models of the dark energy.

\section{Fitting the supernovae data with a cosmological constant \label{cosconst}}

If $R(t)$ is the scale parameter in the Robertson-Walker metric and $k$ its
curvature constant, the Friedmann equations with a cosmological constant are
\cite{weinberg},
\enlargethispage{.2in}
\begin{eqnarray}
\frac{\dot{R}^2}{R^2}+\frac{k}{R^2} & =
&\frac{8\pi\,G}{3}\rho+\frac{\Lambda}{3}\,,\label{vel} \\
\frac{\ddot{R}}{R} & = &
-\frac{4\pi\,G}{3}(\rho+3p)+\frac{\Lambda}{3}\,,\label{acc}
\end{eqnarray}
where $G$ is Newton's constant, $\rho$ is the density of matter, $p$ its
pressure and a dot denotes time differentiation. Conservation of energy and
momentum gives the additional relation
\begin{equation}\label{cons}
d(\rho\,R^3)=-pd(R^3)\,.
\end{equation}

Letting $H^2=\dot{R}^2/R^2$, and denoting the present time by the subscript 0,
Eq.\,(\ref{vel}) can be divided by $H_0^2$, the square of the Hubble constant,
to give
\begin{equation}
1 =
\frac{1}{H_0^2}\left(\frac{8\pi\,G}{3}\rho_0+\frac{\Lambda}{3}-\frac{k}{R_0^2}\right)
\;\equiv\; \left(\Omega_M+\Omega_{\Lambda}+\Omega_k\right)\,,
\end{equation}
at present and
\begin{equation}\label{hsq}
H^2=H_0^2\left(\Omega_M\frac{\rho}{\rho_0}+\Omega_{\Lambda}+\Omega_k\frac{R_0^2}{R^2}\right)\,.
\end{equation}
in general. Assuming the pressure of ordinary matter is negligible,
Eq.\,(\ref{cons}) gives
\begin{equation}
\frac{\rho}{\rho_0}=\frac{R_0^3}{R^3}\,,
\end{equation}
and, in terms of the scaled variables $y=R/R_0$ and $\tau=H_0(t-t_0)$,
Eq.\,(\ref{hsq}) becomes
\begin{equation}\label{hsqy}
\left(\frac{dy}{d\tau}\right)^2=\frac{1}{y}\left(\Omega_M-(\Omega_M+\Omega_{\Lambda}-1)y+
\Omega_{\Lambda}y^3\right)\,.
\end{equation}
From Eq.\,(\ref{hsqy}), the time history of the universe can be obtained as
\begin{equation}\label{hist}
H_0(t-t_0)=\int_1^y\,dy'\frac{\sqrt{\ds
y'}}{\sqrt{\ds\Omega_M-(\Omega_M+\Omega_{\Lambda}-1)y'+\Omega_{\Lambda}y'^3}}\,.
\end{equation}
Owing to the cubic behavior of the denominator of Eq.\,(\ref{hist}), the
$\Omega_M-\Omega_{\Lambda}$ plane is divided into different categories of
histories depending on whether a zero of the cubic occurs for $y<0$, $0<y<1$,
or $y>1$. These are, respectively, those universes which continue to increase
in size from $y=0$, those which never achieve $y=0$ and those which start at
$y=0$, increase to a maximum size and then collapse. The boundaries of these
regions can be seen, for example, in Fig.\,\ref{p_w=-1} along with the line
corresponding to flat cosmologies, $\Omega_{\Lambda}=1-\Omega_M$. Also, at the
present time, Eq.\,(\ref{acc}) can be written
\begin{equation}
\frac{\ddot{R}_0}{R_0}=H_0^2\left(-\frac{\Omega_M}{2}+\Omega_{\Lambda}\right)\,,
\end{equation}
and hence the universe is accelerating if $\Omega_{\Lambda}>\Omega_M/2$.

The extraction of the cosmological parameters $\Omega_M$ and
$\Omega_{\Lambda}$ is achieved by relating the apparent magnitudes of $m(z)$ of
the type Ia supernovae and their absolute magnitudes $M$ to their luminosity
distance $d_L(z)$ as
\begin{equation}
m(z)-M=5\log_{10}\left(d_L(z)\right)+25\,,
\end{equation}
where $d_L$ is measured in megaparsecs. The luminosity distance of a supernova
at red shift $z$ can be expressed as
\begin{equation}\label{dl}
d_L(z)=\frac{c(1+z)}{H_0\sqrt{\ds|\Omega_k|}}\left\{\begin{array}{c}
                                                \sin \\
                                                1    \\
                                                \sinh
                                                \end{array}\right\}\left[\sqrt{|\Omega_k|}
\int_0^z\,dz'\frac{1}{\sqrt{\ds
(1+z')^2(1+z'\Omega_M)-z'(2+z')\Omega_{\Lambda}}}\right]\,,
\end{equation}
with $c$ denoting the velocity of light. In Eq.\,(\ref{dl}), $\sin[\;]$ is used
for $\Omega_k > 0$, $\sinh[\;]$ is used for $\Omega_k < 0$, and the unmodified square bracket is used for $\Omega_k=0$.

To obtain a fit to the data, we write the apparent magnitude as
\begin{equation}\label{appmag}
m(z,\Omega_M,\Omega_{\Lambda},\mathcal{M})=5\log_{10}(\mathcal{D}_L)+\mathcal{M}\,,
\end{equation}
where $\mathcal{D}_L=c^{-1}H_0d_L$ and $\mathcal{M}=M+5\log_{10}(cH_0^{-1})+25$
is an additive constant. The parameters $\Omega_M$, $ \Omega_{\Lambda}$ and
$\mathcal{M}$ are then determined by minimizing
\begin{equation}\label{chisq}
\chi^2=\sum_i\frac{\left(m_{\rm
exp}(z_i)-m(z_i,\Omega_M,\Omega_{\Lambda},\mathcal{M})\right)^2}{\sigma_i^2}\,,
\end{equation}
with $\sigma_i$ denoting the error in $m_{\rm exp}(z_i)$.

Our fit \cite{jpc} to the data presented in Ref.\,\cite{super} is shown in
Fig.\,\ref{p_w=-1}. To avoid repetition, we restrict our fits to this data. We have obtained similar results for the data in Ref.\,\cite{hiz}. When computing points on the $\Delta\chi^2=\chi^2-\chi_{\rm min}^2$ contours, we choose the minimum
value of $\mathcal{M}$ at each point ($\Omega_M$,$\Omega_{\Lambda}$). This is
easily done because the minimization condition for $\mathcal{M}$ can be solved
explicitly in terms of $\Omega_M$ and $\Omega_{\Lambda}$. The values of
$\Delta\chi^2$ plotted (2.30,4.61,6.18) would correspond to 67.3\%, 90.0\% and
95.5\% confidence contours if the errors were Gaussian. The contours are not
precisely elliptical indicating non-Gaussian behavior, and this is addressed in
the detailed analysis of Refs.\,\cite{hiz} and \cite{super}. Fortunately, our
more naive analysis yields very similar $\Delta\chi^2$ contours, which we can
use to assess the quality of the fits resulting from different assumptions
about the dark component.

For the most part, the fit shown in Fig.\,\ref{p_w=-1} lies
above the dashed line separating accelerating from non-accelerating
cosmologies. While the best fit is not a flat cosmology, the $\chi^2$ values
in Table \ref{fits} show that requiring $\Omega_M+\Omega_{\Lambda}=1$ does not
dramatically alter the quality of the fit.

\section{Varying the equation of state \label{quin}}

To generalize beyond the case of a constant dark energy, we can write
Eq.\,(\ref{vel}) as
\begin{equation}\label{velx}
\frac{\dot{R}^2}{R^2}=H_0^2\left(\Omega_M\frac{\rho_M}{\rho_{M0}}+\Omega_X
\frac{\rho_X}{\rho_{X0}}+\Omega_k\frac{R_0^2}{R^2}\right)\,,
\end{equation}
and determine the dark energy density $\rho_X$ using the energy conservation
equation
\begin{equation}
d\rho_X=-3\left(\rho_X+p_X\right)\frac{dR}{R}\,,
\end{equation}
together with the equation of state
\begin{equation}
p_X=w(R)\,\rho_X\,.
\end{equation}
For models of this type, the generalization of Eq.\,(\ref{hsqy}) is
\begin{equation}
\left(\frac{dy}{d\tau}\right)^2=\frac{1}{y}\left(\Omega_M-(\Omega_M+\Omega_X-1)y+
\Omega_Xe^{-3\int_1^y\frac{dy'}{y'}w(y')}\right)\,,
\end{equation}
where $\Omega_X$ is
\begin{equation}\label{omx}
\Omega_X=\frac{8\pi\,G}{3H_0^2}\rho_{X0}\,.
\end{equation}

When $w(y)$ is a constant, say $w=-a,\,a>0$, the boundaries in the
$\Omega_M-\Omega_X$ plane between regions of expanding universes and universes
which eventually collapse or have no big bang are obtained by determining
where the minimum of the function
\begin{equation}
\Omega_M-(\Omega_M+\Omega_X-1)y+\Omega_Xy^{3a}
\end{equation}
vanishes. This leads to the equation
\begin{equation}\label{bndry}
\frac{\Omega_X}{\Omega_M}=\frac{(3a-1)^{(3a-1)}}{(3a)^{3a}}\left(\frac{\Omega_X}{\Omega_M}+
\frac{(\Omega_M-1)}{\Omega_M}\right)^{3a}\,,
\end{equation}
from which $\Omega_X$ can be found as a function of $\Omega_M$. When $a=1$,
Eq.\,(\ref{bndry}) gives the familiar results for the case of a cosmological
constant \cite{carroll}. For these models, the generalization of
Eq.\,(\ref{acc}) is
\begin{equation}\label{quinacc}
\frac{\ddot{R}}{R}=-\frac{4\pi\,G}{3}\rho_M-\frac{4\pi\,G}{3}(1-3a)\rho_X\,.
\end{equation}
At the present time, this gives the acceleration parameter
\begin{equation}
\frac{\ddot{R}_0}{R_0}=\left(-\Omega_M+(3a-1)\Omega_X\right)\frac{H_0^2}{2}\,,
\end{equation}
and hence the universe is accelerating if
\begin{equation}
\Omega_X>\frac{\Omega_M}{(3a-1)}\,.
\end{equation}

\begin{figure}[h]
\centering\includegraphics[height=3.0in]{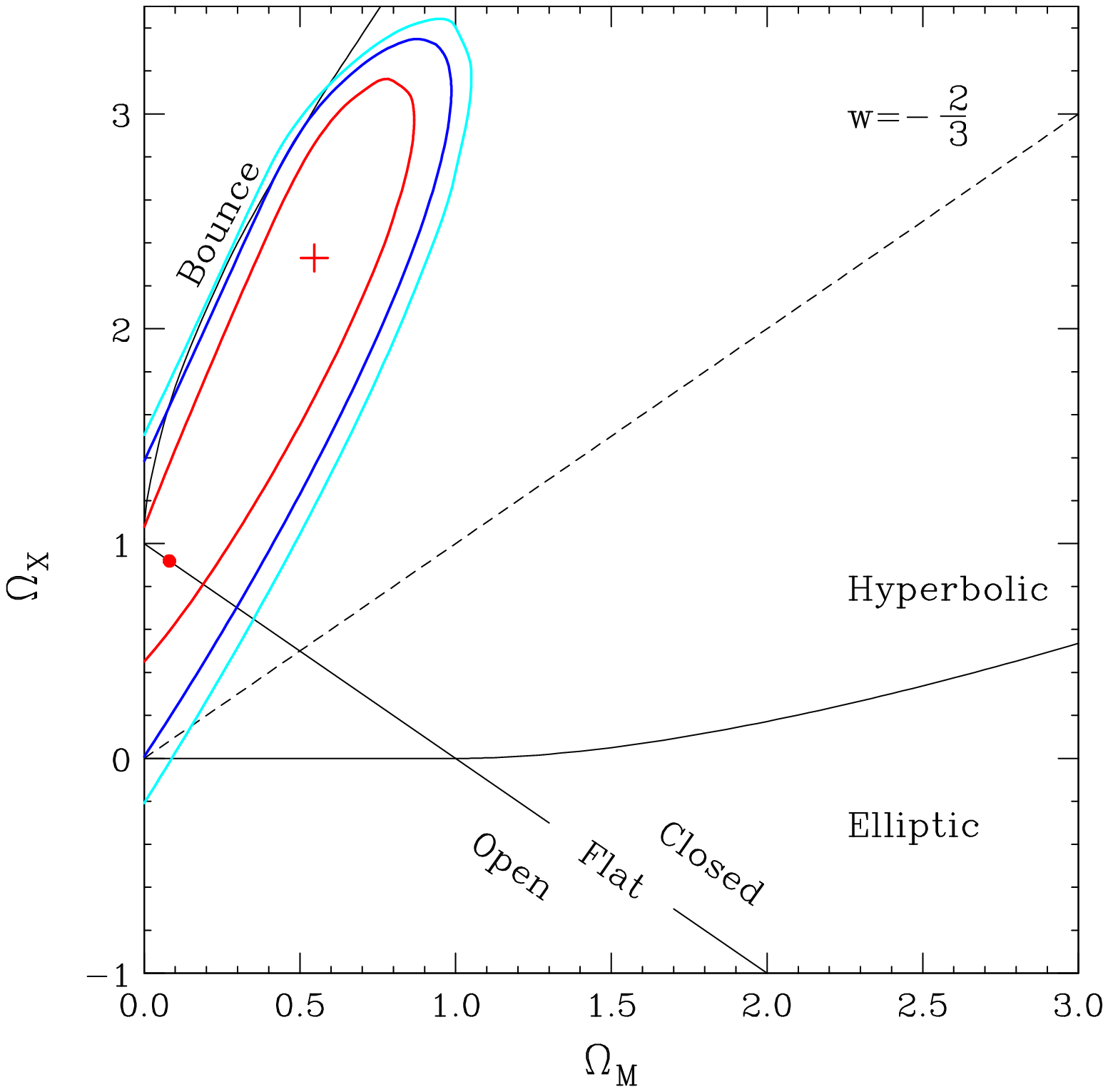}
\caption{\footnotesize The $\Delta\chi^2$ contours for the fit to the 54
supernovae analyzed in Ref.\,\cite{super}, assuming a constant equation of
state of the form $w=-2/3$, are shown. The contours correspond to the 68.3\%,
90\% and 95.4\% confidence levels. The cross indicates the location of the
minimum and the dot indicates the location of the minimum assuming a flat
cosmology. The dashed line separates accelerating from non-accelerating
cosmologies. \label{p_w=-067}}
\end{figure}

\begin{figure}[h]
\centering\includegraphics[height=3.0in]{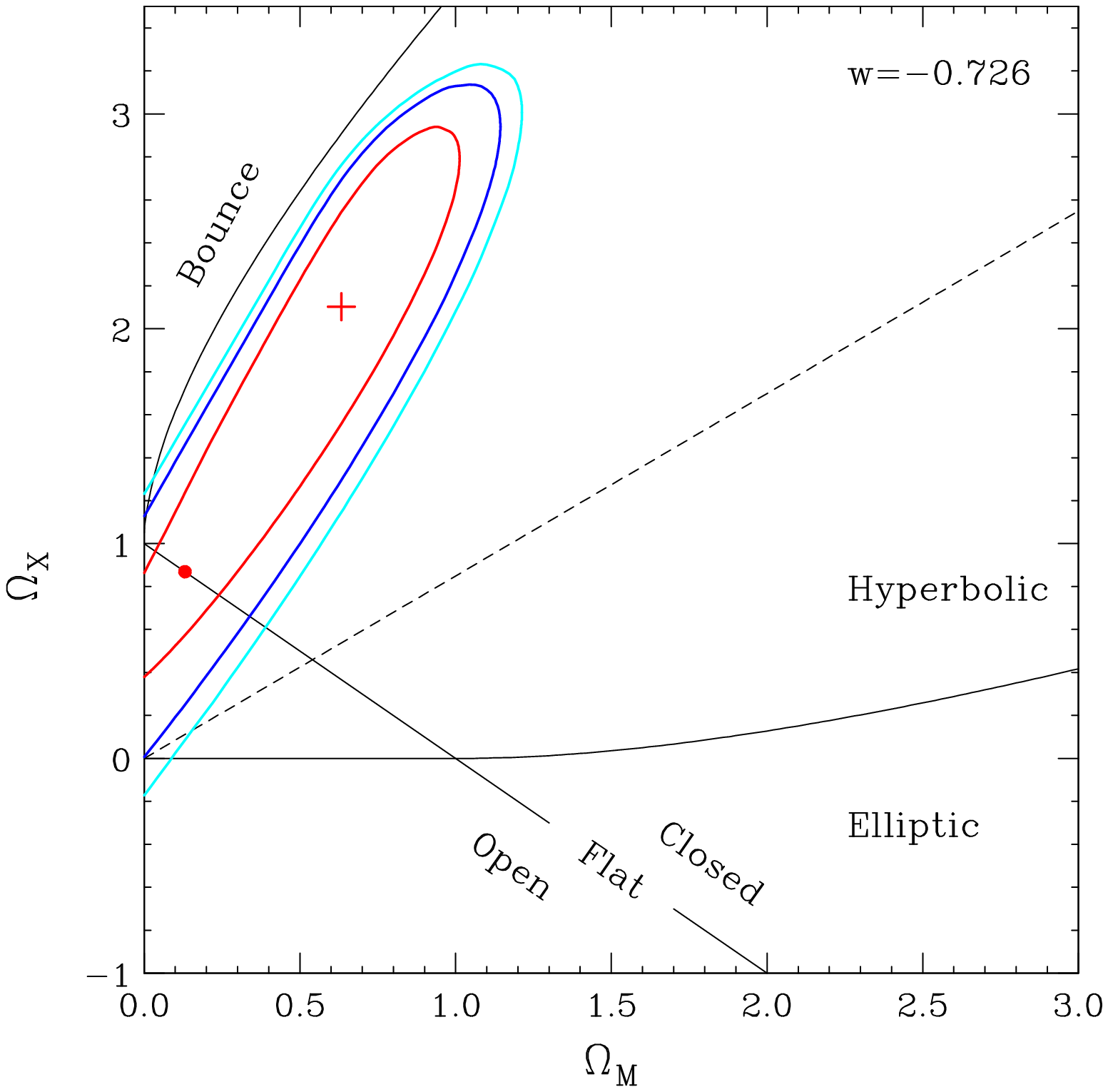}
\caption{\footnotesize The $\Delta\chi^2$ contours for the best fit to the 54
supernovae analyzed in Ref.\,\cite{super}, assuming a constant equation of
state of the form $w=-a$, are shown. The contours correspond to the 68.3\%,
90\% and 95.4\% confidence levels. The cross indicates the location of the
minimum and the dot indicates the location of the minimum assuming a flat
cosmology. The dashed line separates accelerating from non-accelerating
cosmologies. \label{p_w=-073}}
\end{figure}

To gain some sense of what varying the value of $a$ does to the quality of the
fits, we performed a $\chi^2$ analysis of the data in Ref.\,\cite{super} for
the case $a=-2/3$ and for the case where $a$ is varied along with the other
parameters to obtain the best fit. The results are shown in
Figs.\,\ref{p_w=-067} and \ref{p_w=-073} and the best fits are given in Table
\ref{quin1}. These examples show that the quality of the fits to the data of
Ref.\,\cite{super} for the cases considered -- $w=-1,\,-2/3\,,-0.73$ -- is
basically the same. Even when the value of $w$ is allowed to vary, the minimum
in $\chi^2$ is not lowered in any significant way. One may argue that values
of $w\geq -2/3$ are less preferred since, as suggested by
Fig.\,\ref{p_w=-067}, their $\chi^2$ contours encroach into the region of
cosmologies which have no big bang.

\begin{figure}[h]
\centering\includegraphics[height=3.0in]{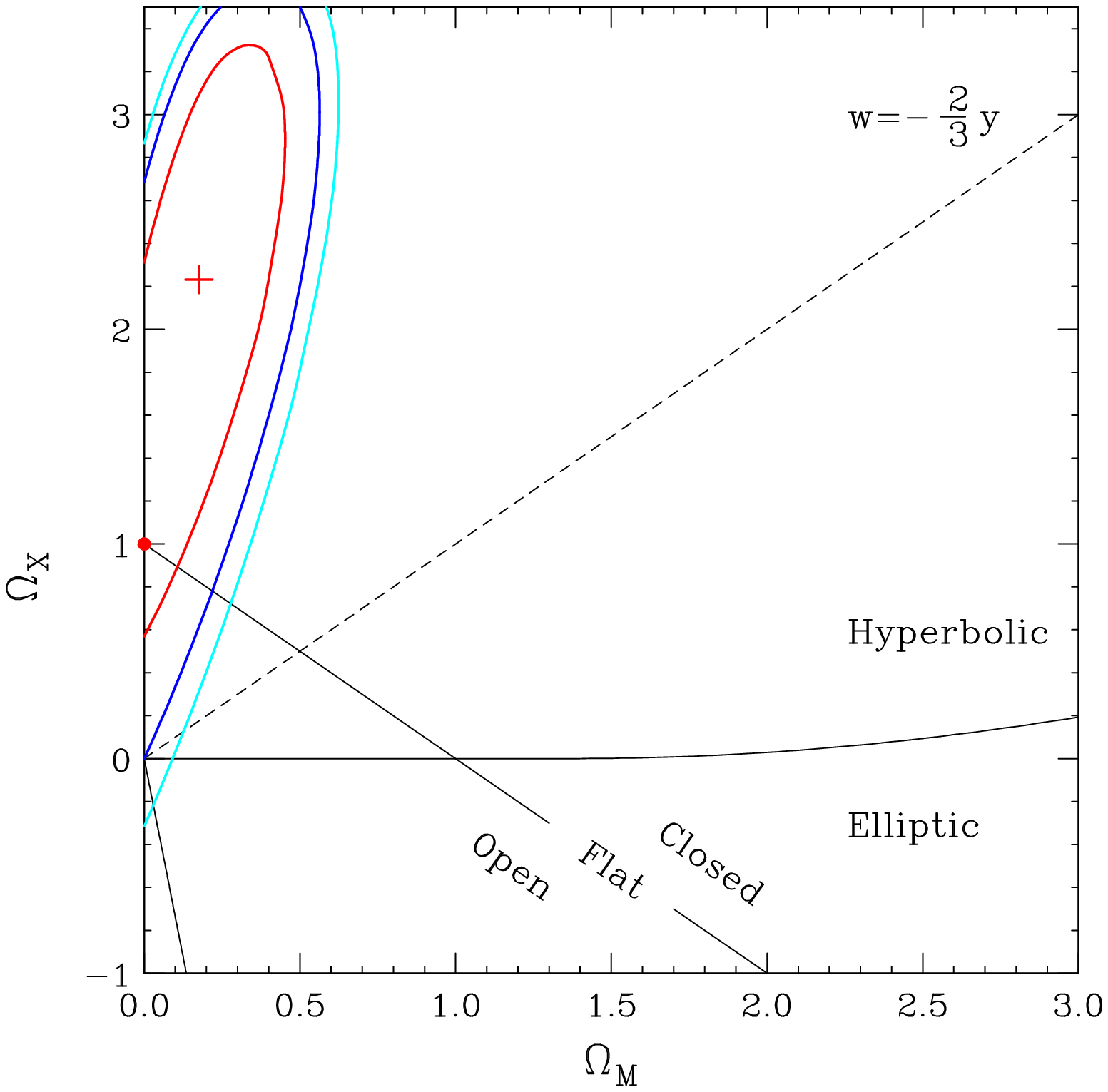}
\caption{\footnotesize The $\Delta\chi^2$ contours for the best fit to the 54
supernovae analyzed in Ref.\,\cite{super}, assuming an equation of state of the
form $w=-2y/3$, are shown. The contours correspond to the 68.3\%, 90\% and
95.4\% confidence levels. The cross indicates the location of the minimum and
the dot indicates the location of the minimum assuming a flat cosmology. The
dashed line separates accelerating from non-accelerating cosmologies.
\label{p_w=-067y}}
\end{figure}

The constant equations of state discussed above ($w=-a$) all result power law
dark energy densities of the form
\begin{equation}\label{power}
\frac{\rho_X}{\rho_{X0}}=y^{3(a-1)}\,,
\end{equation}
and all lead to fits with comparable $\chi^2$. To determine how changing the
functional dependence of $\rho_X$ on $y$ affects the quality of the fits, we
examined equations of state of the form $w=-by,\,b>0$. These equations of
state lead to dark energy densities with an exponential dependence of the form
\begin{equation}
\frac{\rho_X}{\rho_{X0}}=\frac{e^{3b(y-1)}}{y^3}\,.
\end{equation}
In order to make direct comparisons with the constant $w$ cases, we fit the
data of Ref.\,\cite{super} using $b=2/3$ and $b=1$. These fits are shown in
Figs.\,\ref{p_w=-067y} and \ref{p_w=-y}. The numerical details are given in
Table\,\ref{quin2}. Again, the $\chi^2$ of these fits indicates that data are
not too sensitive to the $y$-dependence of $w$. However, the location and size
of the contours suggest that the $w=-y$ choice is the more desirable.

\begin{figure}[h]
\centering\includegraphics[height=5.0in]{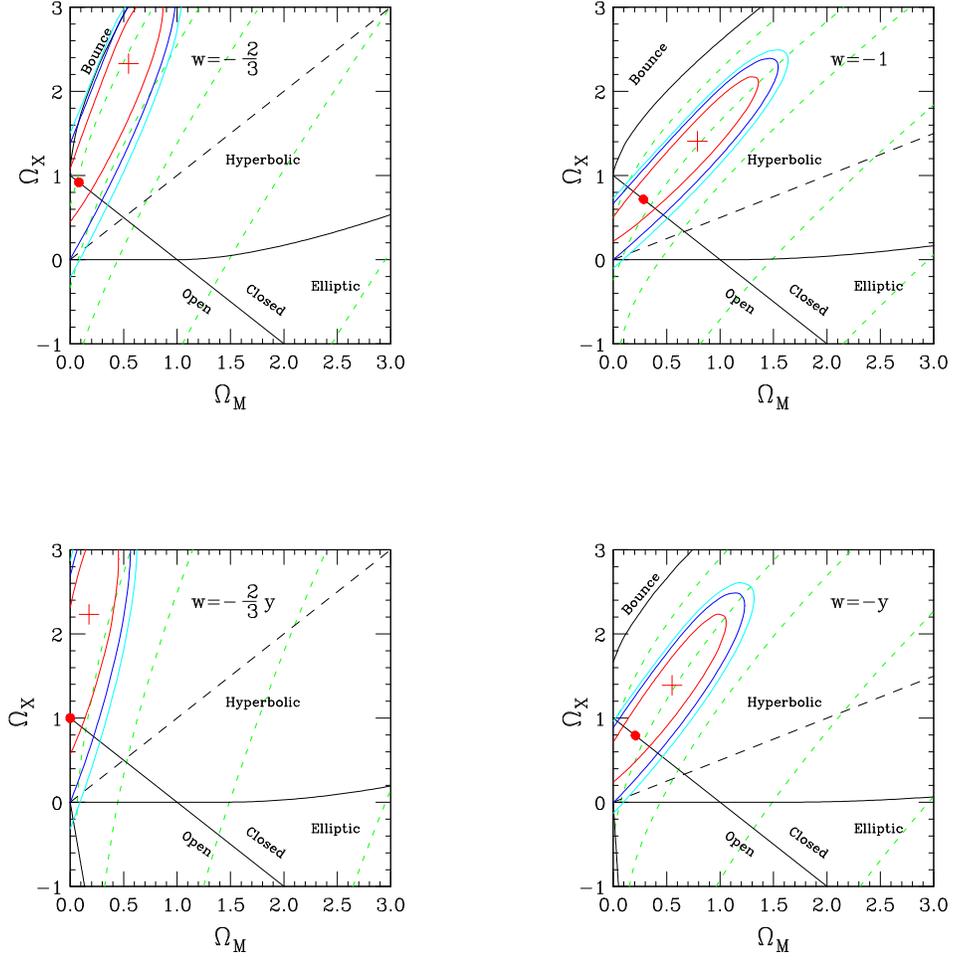}
\caption{\footnotesize The $\chi^2$ contours and (dashed) lines of constant lifetime are shown. Starting from the lower right and moving to the upper left of each panel, the dashed green lines correspond to 8.0, 9.5, 11.9, 14.3 and 19.0 Gyr. In computing the lifetimes, we used $H_0$=63 km/sec/Mpc. In the $w=-2y/3$ panel, the 19.0 Gyr line is not visible on the portion of the plane shown.
\label{newlife_2}}
\end{figure}

The lifetime of the universe depends on the choice of $w$ and can be used to
discriminate among various models. Fig.\,\ref{newlife_2} shows the
$\Omega_M-\Omega_X$ plane for four different choices of $w$ along with lines of
constant lifetime. Generally, the $\chi^2$ contours lie between 11.9 and 19.0
Gyr.

\section{Dark energy as a viscous fluid \label{visc}}

As an alternative to simply modifying the equation of state of the dark
component, in this section we consider the consequences of assuming the dark
energy consists of a fluid having a bulk viscosity $\zeta$. If a bulk
viscosity term is introduced into the energy-momentum tensor of an ideal fluid,
one effect is to replace Eq.\,(\ref{cons}) with \cite{weinberg}
\begin{equation}\label{consv}
d(\rho_X\,R^3)=-p_X^*d(R^3)\,\equiv\,-\left(p_X-3\,\zeta\,\frac{\dot{R}}{R}\right)d(R^3)\,.
\end{equation}
In view of Eq.\,(\ref{velx}), the expression for $\dot{R}/R$ depends on the
dark energy density $\rho_X$ in a nonlinear fashion, which means that, unlike
the case of the multiplicative equations considered previously,
Eq.\,(\ref{consv}) is not separable. In general, it must be solved numerically.

If we assume that the pressure $p_X$ of the dark component is negligible, the
differential equation determining $\rho_X$ is
\begin{equation}
d\rho_X=-3\left(\rho_X-3\,\zeta\,\frac{\dot{y}}{y}\right)\frac{dy}{y}\,,
\end{equation}
or, using Eq.\,(\ref{velx}),
\begin{equation}
d\left(\frac{\rho_X}{\rho_{X0}}\right)=-3\left(\frac{\rho_X}{\rho_{X0}}-3\frac{\zeta
H_0}{\rho_{X0}}\sqrt{\ds
\frac{\Omega_M}{y^3}+\Omega_X\frac{\rho_X}{\rho_{X0}}+\frac{\Omega_k}{y^2}}\;\right)\frac{dy}{y}\,.
\end{equation}
It is more convenient to write a differential equation for
\begin{equation}\label{hsqx}
\frac{H^2}{H_0^2}\equiv\mathcal{H}^2=\frac{\Omega_M}{y^3}+\Omega_X\frac{\rho_X}{\rho_{X0}}+\frac{\Omega_k}{y^2}\,,
\end{equation}
since $\mathcal{H}$ is the quantity needed to determine the luminosity distance $d_L$. This differential equation is
\begin{equation}\label{hsqde}
\frac{d\mathcal{H}^2}{dy}=-3\frac{\mathcal{H}^2}{y}+3A\frac{\mathcal{H}}{y}+\frac{\Omega_k}{y^3}\,,
\end{equation}
where $A$ is given by
\begin{equation}\label{adef}
A=\frac{3\Omega_X\zeta H_0}{\rho_{X0}}\,=\,\frac{8\pi\,G\zeta}{H_0}\,.
\end{equation}
The second equality follows from Eq.\,(\ref{omx}). Since the luminosity distance is given by
\begin{equation}\label{viscdl}
d_L(z)=\frac{c\,(1+z)}{H_0\sqrt{\ds |\Omega_k|}}\left\{\begin{array}{c}
                                                \sin \\
                                                1    \\
                                                \sinh
                                                \end{array}\right\}\left[\sqrt{|\Omega_k|}
\int_0^z\frac{dz'}{\mathcal{H}(z')}\right]\,,
\end{equation}
where $\mathcal{H}(z)$ is obtained by replacing $y$ by $1/(1+z)$, we write the
differential equation in terms of $z$, and solve
\begin{equation}\label{dhdz}
\frac{d\mathcal{H}^2(z)}{dz}=3\frac{\mathcal{H}^2(z)-A\mathcal{H}(z)}{(1+z)}-\Omega_k(1+z)\,.
\end{equation}
From the form of Eq.\,(\ref{dhdz}), the expression for the apparent magnitude
$m(z)$ depends on the parameters $A$ and $\Omega_k$, which we can vary to
obtain a fit to the supernova data. We perform the integration in Eq.\,(\ref{viscdl}) numerically, determining $\mathcal{H}(z')$ using a four point Runge-Kutta algorithm and the boundary condition $\mathcal{H}(0)=1$ to first solve the differential equation. Given $d_L(z)$, for the data point $z_i$, we then minimize $\chi^2$ as in Eqs.\,(\ref{appmag}) and (\ref{chisq}). The resulting $\chi^2$ contours are shown in Fig.\,\ref{viscage_1} and the best fits are listed in Table \ref{viscfit}. In Fig.\,\ref{viscage_1}, the lines corresponding to various values of $\Omega_M$ indicate the boundary of the region below which, for that $\Omega_M$, the dark energy density $\rho_X$ remains positive. This is determined by the condition
\begin{equation}\label{rhopos}
\Omega_X\frac{\rho_X}{\rho_{X0}}=\mathcal{H}^2(z)-(1+z)^3\Omega_M-(1+z)^2\Omega_k\geq
0\,,
\end{equation}
which is analyzed by choosing value of $\Omega_M$, fixing $A$, taking $\Omega_k$ to be large and negative and solving Eq.\,(\ref{dhdz}) for $1\leq z\leq 1000$. If condition Eq.\,(\ref{rhopos}) is satisfied for all $z$, $\Omega_k$ is increased and the process is repeated until, for a sufficiently large value of $\Omega_k$, Eq.\,(\ref{rhopos}) is violated for some $z$. This determines a point $(A,\Omega_k)$ on the curve for the given $\Omega_M$ and the region above this curve is not allowed. 

\begin{figure}[h]
\centering\includegraphics[height=3.0in]{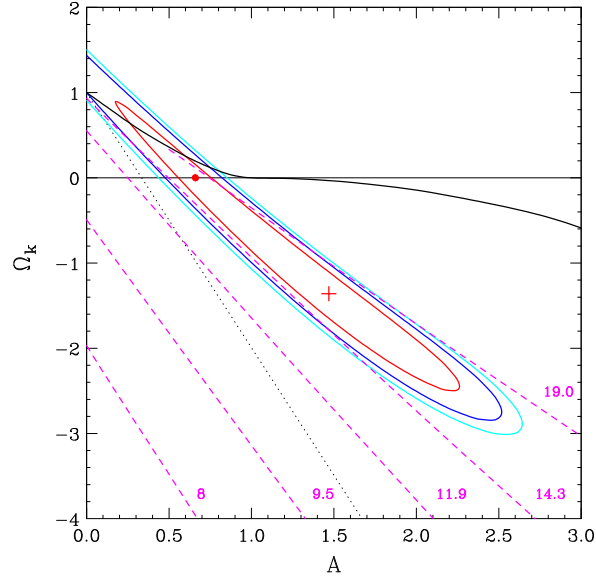}
\caption{\footnotesize The $\Delta\chi^2$ contours for the best fit to the 54
supernovae analyzed in Ref.\,\cite{super}, assuming a viscous medium, are
shown. The contours correspond to the 68.3\%, 90\% and 95.4\% confidence
levels. The cross indicates the location of the minimum $\chi^2$ and the dot indicates the location of the minimum $\chi^2$ assuming a flat cosmology. The dashed lines are the lines of constant lifetime. The region above the dotted line corresponds to accelerating cosmologies. \label{viscage_2}}
\end{figure}

The range of lifetimes for this case, obtained from the relation 
\begin{equation}\label{visclife}
t_0H_0=\int_0^1dy\frac{1}{y\mathcal{H}(y)}\,,
\end{equation}
is shown in Fig.\,\ref{viscage_2}. Here, unlike the cases of Secs.\,\ref{cosconst} and \ref{quin}, the lifetime cannot be much less than 14 Gyr. The lines of constant lifetime terminate on the dark black line in this figure and Fig.\,\ref{viscage_1}. For $\Omega_k$ and $A$ above this line, the lifetime integral does not converge. In each of these figures, the dotted line separates the accelerating from the non-accelerating cosmologies, with the region above the line corresponding to accelerating cosmologies. This can be seen by noting that analog of Eq.\,(\ref{quinacc}) is
\begin{equation}
\frac{\ddot{R}}{R}=-\frac{4\pi\,G}{3}\rho_{M}-\frac{4\pi\,G}{3}\left(\rho_{X}+3p^*_X\right)\,,
\end{equation}
which gives  at the present time
\begin{eqnarray}
\frac{\ddot{R_0}}{R_0}& = &
-\frac{4\pi\,G}{3}\left(\rho_{M0}+\rho_{X0}\right)+12\pi\,G\zeta H_0 \nonumber \\
& = & \frac{H_0^2}{2}\left(3A+\Omega_k-1\right)\,.
\end{eqnarray}
Hence, accelerating cosmologies satisfy $\Omega_k\,>\,-3A+1$.

When $\rho$ and $p$ are related by an equation of state of the form
$p=w\,\rho$, the cosmology evolves at constant entropy. In the present case,
there is entropy production. This can be seen by noting that \cite{weinberg_1}
\begin{equation}
\frac{dS}{dt}=\frac{2\pi^2}{T(R)}\left(\frac{d(\rho_XR^3)}{dt}+p_X\frac{dR^3}{dt}\right)
\;=\;\frac{2\pi^2}{T(R)}\left(\frac{d(\rho_XR^3)}{dt}+\left(p_X^*+3\zeta\frac{\dot{R}}{R}\right)
\frac{dR^3}{dt}\right)\,,
\end{equation}
where $T(R)$ is the temperature of the dark energy. On using Eq.\,(\ref{consv}), we have
\begin{equation}
\frac{dS}{dt}=9\zeta\frac{2\pi^2R^3}{T(R)}\left(\frac{\dot{R}^2}{R^2}\right)\,.
\end{equation}
Considering $S$ as a function of $y=R/R_0$, the evolution of $S(y)$ is given by
\begin{equation}
\frac{dS}{dy}=\frac{9\zeta V_0}{T(y)}\left(\frac{\dot{y}}{y}\right)y^2\,,
\end{equation}
where $V_0=2\pi^2R_0^3$. With the aid of Eq.\,(\ref{hsqx}), the entropy is
\begin{equation}\label{deltaS}
S(y)-S_0=9\zeta V_0H_0\int_1^y\,dy'\frac{y'^2\mathcal{H}(y')}{T(y')}\,,
\end{equation}
where $\mathcal{H}(y)$ is obtained by solving Eq.\,(\ref{hsqde}). Setting
\begin{equation}\label{hsqu}
\mathcal{H}^2(y)=y^{-3}u(y)
\end{equation}
and using Eq.\,(\ref{adef}) to eliminate $\zeta$, the change in the entropy density at present is given by
\begin{equation}\label{entden}
\frac{\Delta S_0}{V_0}=\frac{9AH_0^2}{8\pi\,G}
\int_0^1\,dy\frac{\sqrt{\sb y}\sqrt{u(y)}}{T(y)}
\;=\;3A\rho_{\rm crit}\int_0^1\,dy\frac{\sqrt{\sb y}\sqrt{u(y)}}{T(y)}\,,
\end{equation}
where $\rho_{\rm crit}=3H_0^2/8\pi\,G$ is the critical density and $u(y)$
satisfies
\begin{equation}\label{dudy}
\frac{du}{dy}=3A\sqrt{\sb y}\sqrt{u}+\Omega_k\,.
\end{equation}
To complete the calculation of $\Delta S_0$, the functional form of $T(y)$ is needed. The choice of $T(y)$ must be consistent with the integrability of the entropy, which, for $T$ taken to be a function of $V$ and $\rho_X$, requires \cite{Maar}
\begin{equation}\label{integ}
-V\left(\frac{\partial T}{\partial V}\right)_{\!\rho_X}+(\rho_X+p_X)\left( \frac{\partial T}{\partial \rho_X}\right)_{\! V}=T\left( \frac{\partial p_X}{\partial \rho_X}\right)_{\! V}\,.
\end{equation}
Using Eq.\,(\ref{integ}) and Eq.\,(\ref{consv}), it is possible to obtain the relation
\begin{equation}\label{dT}
\frac{dT}{T}=-3\left(\frac{\partial p_X}{\partial \rho_X}\right)_{\! V}\frac{dR}{R} + 9\frac{\zeta H}{T}\left(\frac{\partial T}{\partial \rho_X}\right)_{\! V}\frac{dR}{R}\,.
\end{equation}
Assuming $p_X=0$, Eq.\,(\ref{integ}) implies that $T(V,\rho_X)$ has the general form
\begin{equation}\label{genT}
T(V,\rho_X)=T(V\rho_X)=\int\,d\beta A(\beta)(V\rho_X)^{\beta}\,,
\end{equation}
and, for the purpose of assessing the entropy growth, we examine a single term of the form
\begin{equation}\label{specT}
T=C(V\rho_X)^{\beta}\,,
\end{equation}
where $C$ is a constant. In this case, Eq.\,(\ref{dT}) becomes
\begin{equation}\label{dTy}
\frac{dT}{T}=\beta\,\frac{9\zeta H}{\rho_X}\frac{dR}{R}= \beta\,\frac{3A\sqrt{\sb y}\sqrt{u(y)}\,dy}{u(y)-\Omega_M-y\Omega_k} = \beta\frac{d\!\left[u(y)-\Omega_M-y\Omega_k\right]}{\left[\,u(y)-\Omega_M-y\Omega_k\right]}\,,
\end{equation}
where Eqs.\,(\ref{hsqx}), (\ref{hsqu}) and (\ref{dudy}) have been used. Integrating Eq.\,(\ref{dTy}) gives
\begin{equation}\label{T(y)}
\frac{T(y)}{T(1)}=\frac{\left[u(y)-\Omega_M-y\Omega_k\right]^{\beta}}{\Omega_X^{\beta}} =\frac{(V\rho_X)^{\beta}}{(V_0\rho_{X0})^{\beta}}\,,
\end{equation}
which is consistent with Eq.\,(\ref{specT}). Using Eq.\,(\ref{T(y)}), Eq.\,(\ref{entden}) is also integrable and the entropy density at present is
\begin{equation}
\frac{S_0}{V_0}=\frac{\rho_C\Omega_X}{T(1)}\frac{1}{1-\beta}\left[1 - \left(\frac{u(0)-\Omega_M}{\Omega_X}\right)^{1-\beta}\right]\,.
\end{equation}

As with the luminosity distance above, we evaluate $u(0)$ numerically using a simple four-point algorithm to integrate Eq.\,(\ref{dudy}). The ratio of the present value of the entropy density produced by the viscous medium to the entropy density of the microwave background is plotted in Fig.\,\ref{entropy} for several values of $\beta$ and $T(1)$ taken to be the temperature of the microwave background, $T_{\gamma}=2.73$\,K. The numerical results scale as $T_{\gamma}/T(1)$. Clearly, a side effect of the viscosity being adequate to produce an accelerating cosmology is the production of large amounts of entropy.

\begin{figure}[h]
\centering\includegraphics[height=3.75in]{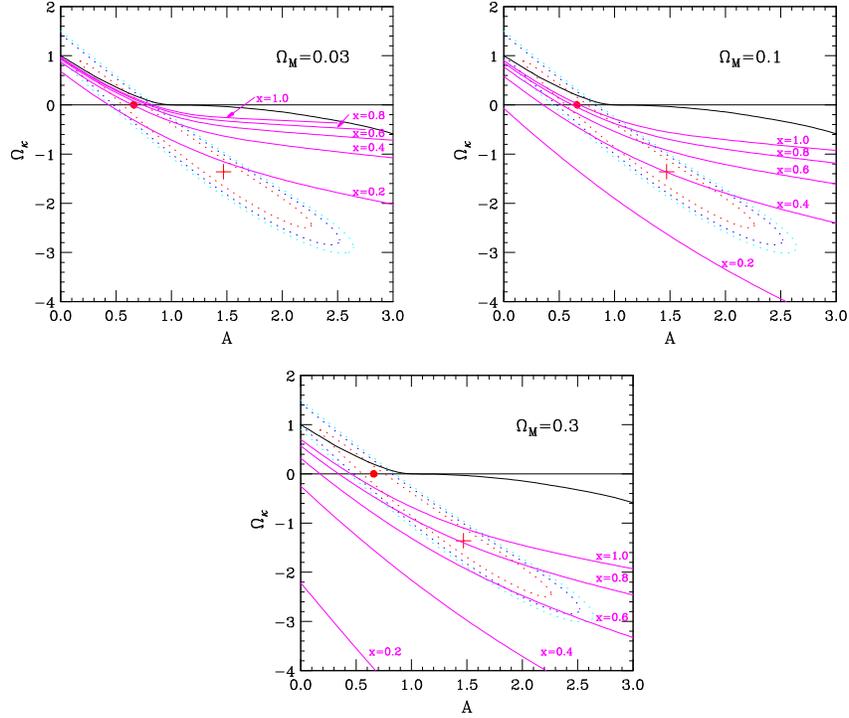}
\caption{\footnotesize Curves of constant values of the exponent $x$ which controls
the growth of the matter density fluctuations $\delta$, $\delta=1/(1+z)^x$, are
shown for several values of $\Omega_M$. \label{viscdel}}
\end{figure}

   Although we are assuming that the matter and dark energy densities
constitute non-interacting components of the total energy density, it is
nevertheless necessary to check the compatibility of this assumption
with the existence of matter density fluctuations which grow at large z
(or small y). For an inviscid fluid in the Newtonian limit, a matter
density fluctuation $\delta$ satisfies the equation
\begin{equation}
\ddot{\delta}+2\frac{\dot{R}}{R}\,\dot{\delta}-4\pi\,G\rho_M\delta=0\,,
\end{equation}
and the influence of the viscous dark energy enters through the
$\dot{R}/R$ term. Using Eqs.\,(\ref{velx}) and (\ref{hsqx}), the differential equation
Eq.\,(\ref{dhdz}), and assuming that $\rho_M\propto (1+z)^3$, the equation for
$\delta$ can be written
\begin{equation}\label{delde}
(1+z)\mathcal{H}^2(z)\,\delta''+\frac{1}{2}\left(\mathcal{H}^2(z)-3A\mathcal{H}(z)-
\Omega_k(1+z)^2\right)\delta'-
\frac{3}{2}\Omega_M(1+z)^2\delta=0\,,
\end{equation}
where the $^{\prime}$ denotes a derivative with respect to $z$. To solve this
equation for large $z$, we note that Eqs.\,(\ref{hsqu}) and (\ref{dudy}) imply
\begin{equation}
\lim_{y\rightarrow 0}\mathcal{H}^2(y)\sim y^{-3}\left(u(0)+\Omega_k\,y\right)\,.
\end{equation}
This gives for large $z$
\begin{equation}\label{hsqz}
\mathcal{H}^2(z)\sim (1+z)^3u(0)+(1+z)^2\,\Omega_k\,.
\end{equation}
Keeping only the leading terms for large $z$, Eq.\,(\ref{delde}) becomes
\begin{equation}
2(1+z)^2\delta''+(1+z)\delta'-3\frac{\Omega_M}{u(0)}\delta=0\,,
\end{equation}
and a solution of the form $\delta=(1+z)^a$ gives
\begin{equation}
a_{\pm}=\frac{1}{4}\left(1\pm\sqrt{\ds 1+24\frac{\Omega_M}{u(0)}}\right)\,.
\end{equation}
Hence, the solution which grows as $z$ decreases is
\begin{equation}
\delta=\left(\frac{1}{1+z}\right)^{x},\quad x=\frac{1}{4}\left(\sqrt{\ds
1+24\frac{\Omega_M}{u(0)}}-1\right)\,.
\end{equation}
In general, $x\leq 1$ because the condition that $\rho_X$ be positive, Eq.\,(\ref{rhopos}), implies, in view of Eq.\,(\ref{hsqz}), that $u(0)\geq \Omega_M$. The curves of constant exponent $x$ for several values of $\Omega_M$ are shown in Fig.\,\ref{viscdel}. The regions below the $x=1$ curves have matter density fluctuations which grow, the trend being toward slower growth for larger negative values of $\Omega_k$.

\section{Conclusions}

When analyzed in terms of a universe consisting of matter and dark energy in the form of a cosmological constant, the supernova data of Refs.\,\cite{hiz,super} favor an accelerating universe. The best fits in this case do not correspond to a flat cosmology, but a flat solution is well within the likelihood contours with a $\chi^2$ relatively close to the minimum value of $\chi^2$. Our ability to reproduce the results of Refs.\,\cite{super} is indicated in Table\,\ref{fits}.

\begin{figure}[h]
\centering\includegraphics[height=3.0in]{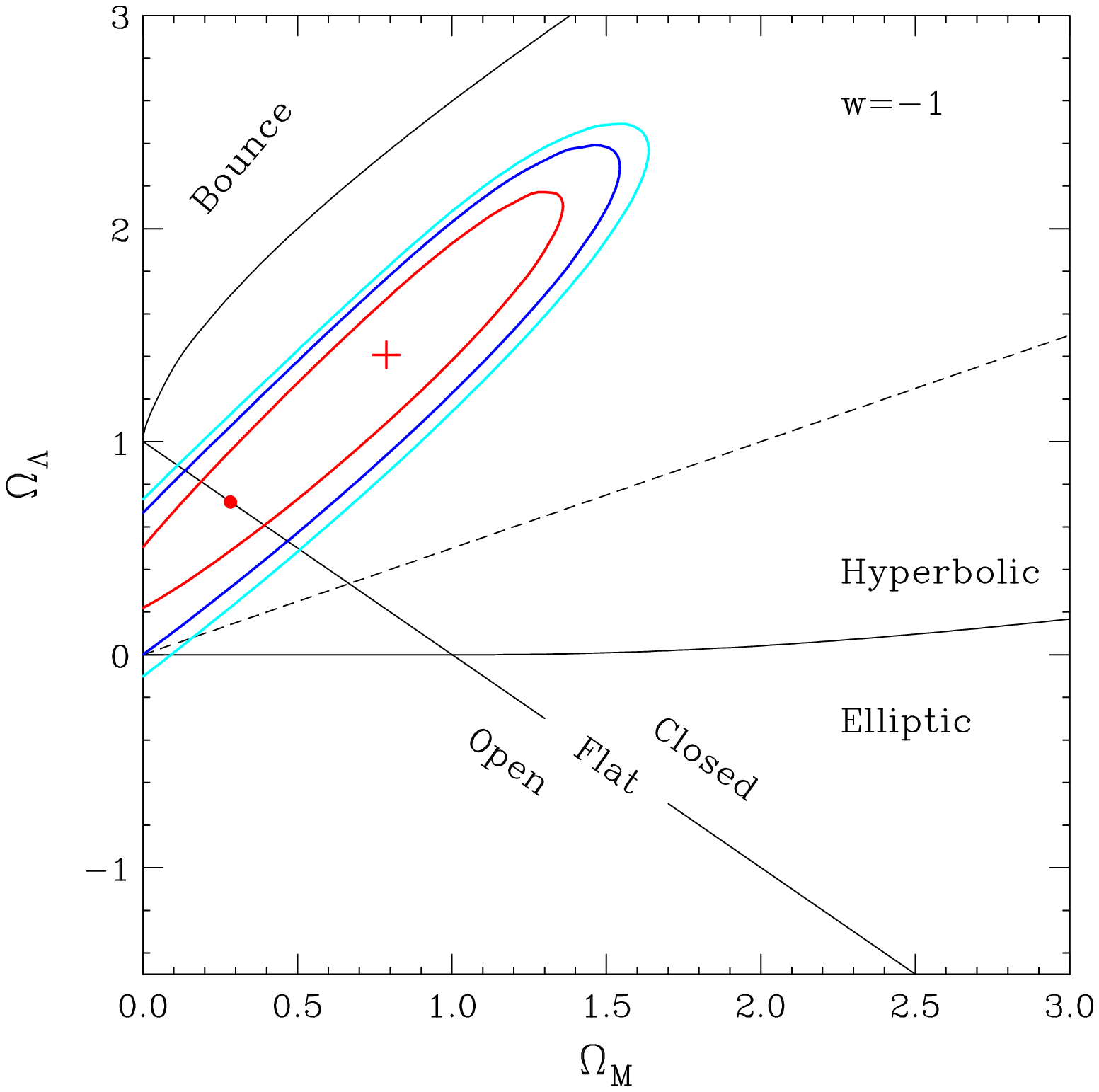}
\caption{\footnotesize The $\Delta\chi^2$ contours for the fit to the 54
supernovae analyzed in Ref.\,\cite{super} are shown. The contours correspond
to the 68.3\%, 90\% and 95.4\% confidence levels. The cross indicates the
location of the minimum and the dot indicates the location of the minimum
assuming a flat cosmology. The dashed line separates accelerating from
non-accelerating cosmologies. \label{p_w=-1}}
\end{figure}

\begin{figure}[h]
\centering\includegraphics[height=3.0in]{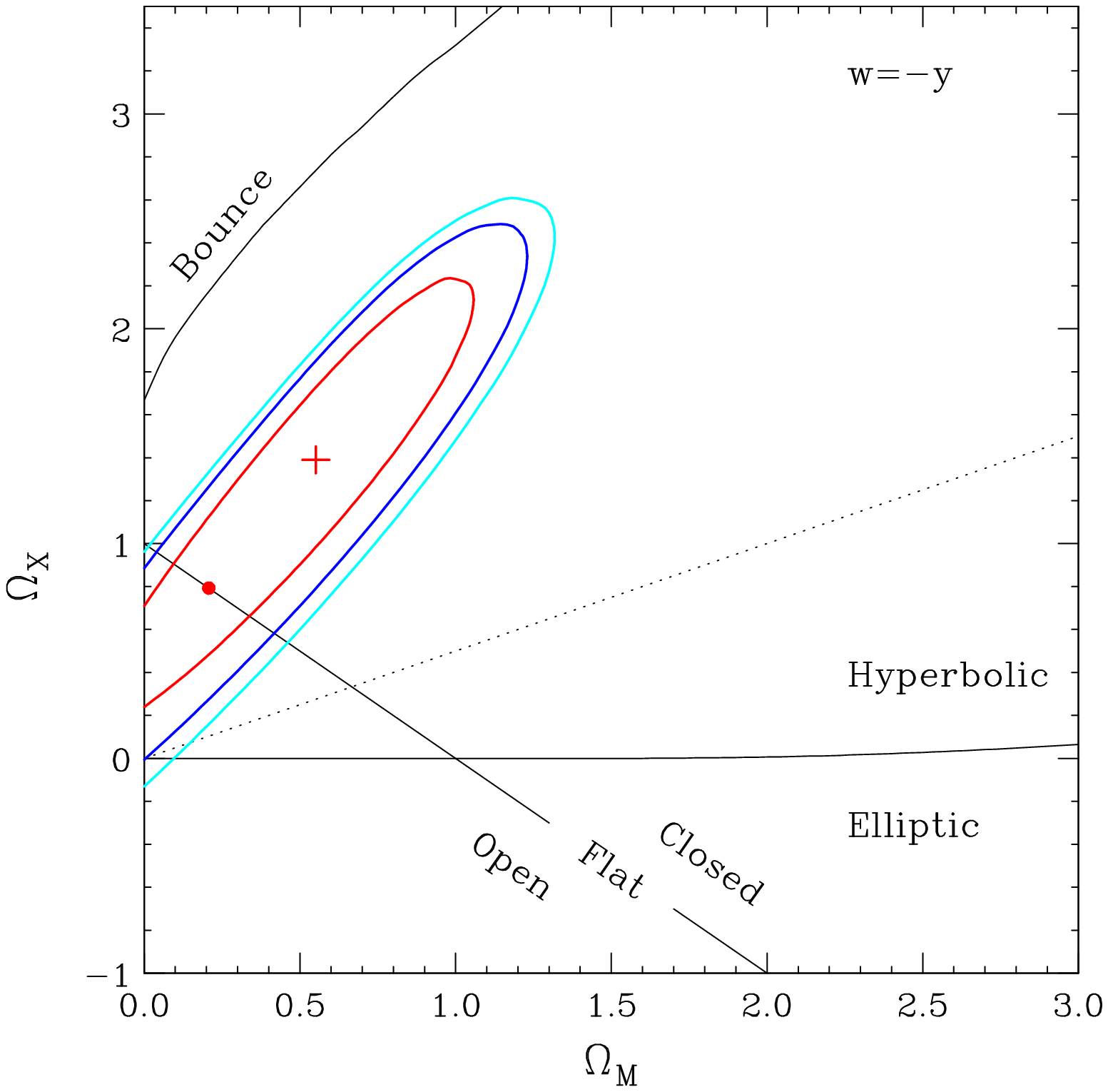}
\caption{\footnotesize The $\Delta\chi^2$ contours for the best fit to the 54
supernovae analyzed in Ref.\,\cite{super}, assuming an equation of state of the
form $w=-y$, are shown. The contours correspond to the 68.3\%, 90\% and 95.4\%
confidence levels. The cross indicates the location of the minimum and the dot
indicates the location of the minimum assuming a flat cosmology. The dashed
line separates accelerating from non-accelerating cosmologies. \label{p_w=-y}}
\end{figure}

As a means of assessing the sensitivity of the currently available supernova data to the form of the equation of state $p=w(y)\rho$, we repeated the analysis using equations of state of the form 
\begin{equation}
w(y)=-a-by\quad\quad a,b>0\,.
\end{equation}
For $b=0$ and a number of different values of $a$, with $0<a<1$, we find best fits with $\chi^2$ very nearly identical to those obtained for a cosmological constant. This includes a determination of the best value of $a$ by treating $a$ as one of the search parameters. In these cases, the sum $\Omega_M + \Omega_X$ tends to be a bit larger than $\Omega_M + \Omega_{\Lambda}$, and the flat cosmology value of $\Omega_M$ is somewhat smaller than the corresponding value obtain for a cosmological constant. When $a=0$ and $b$ is varied, the quality of the fits in terms of $\chi^2$ is again virtually unchanged. The case $w=-y$ is very similar to $w=-1$ in all respects, as can be seen by comparing Figs.\,\ref{p_w=-1} and \ref{p_w=-y}. This tendency of the minimum value of $\chi^2$ to be the same as long as the equation of state describes a dark energy with negative pressure is another indication that the precise form of $w$ is ill determined  by the current supernova data \cite{fhlt}.

To examine whether the introduction of a negative pressure component by means of a multiplicative equation of state was essential, we explored the notion that the dark energy is a fluid with bulk viscosity. The bulk viscosity provides a negative pressure and, again, a fit to the supernova data gives a $\chi^2$ that is indistinguishable from the cosmological constant case (compare Tables \ref{viscfit} and \ref{fits}). We have shown that such a dark energy can provide a fit which is consistent with an accelerating flat cosmology having reasonable values for the lifetime and $\Omega_M$ and density fluctuations with the appropriate behavior. There remains a question of whether the use of a simple linear bulk viscosity term $\zeta\dot{y}/y$ is valid throughout the range of $y(t)$'s probed by the supernova data. There are known non-linear effects, usually associated with viscosity-driven inflation models, which could modify our results \cite{mm,cj,mh}. We have not attempted to include them in this investigation. The model presented here is different from those which suggest that viscosity effects associated with cold dark matter could be the source of the observed acceleration \cite{schwarz,zsbp}. Viscosity effects have also been discussed in models with matter creation \cite{FdSW}. 

\begin{figure}[h]
\centering\includegraphics[height=3.0in]{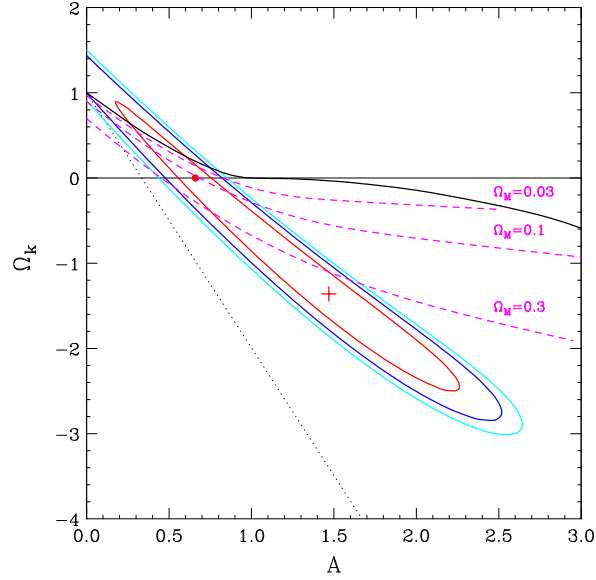}
\caption{\footnotesize The $\Delta\chi^2$ contours for the best fit to the 54
supernovae analyzed in Ref.\,\cite{super}, assuming a viscous medium, are
shown. The contours correspond to the 68.3\%, 90\% and 95.4\% confidence
levels. The cross indicates the location of the minimum $\chi^2$ and the dot indicates the location of the minimum $\chi^2$ assuming a flat cosmology. The dashed lines are the boundaries of the region below which for the given value of $\Omega_M$ the dark energy density $\rho_X$ in positive. The region above the dotted line corresponds to accelerating cosmologies. \label{viscage_1}}
\end{figure}

\begin{figure}[h]
\centering\includegraphics[height=3.0in]{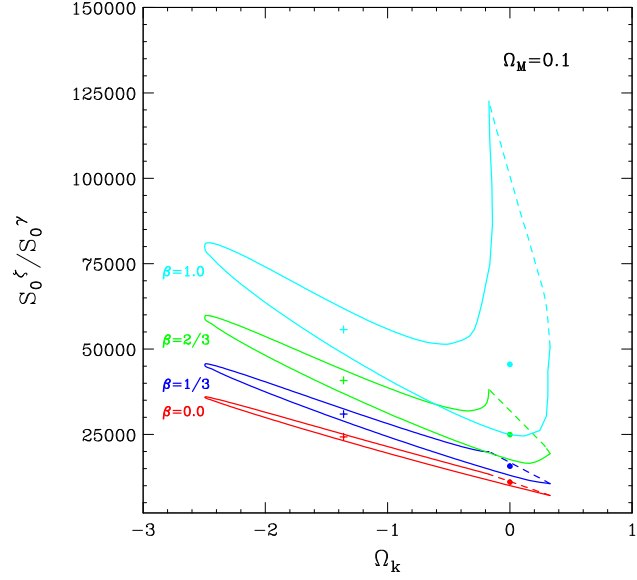}
\caption{\footnotesize The ratio of the present entropy associated with the
viscous medium to the present entropy of the microwave background is plotted
for the 68\% $\Delta\chi^2$  contour corresponding to the best fit to the 54
supernovae analyzed in Ref.\,\cite{super} and $\Omega_M=0.1$. Results for the indicated values of the parameter $\beta$ are shown. The cross indicates the location of the minimum $\chi^2$ and the dot indicates the location of the minimum $\chi^2$ assuming a flat cosmology. The dashed portions of the curves correspond to the $\Omega_M=0.1$ boundary in Fig.\,\ref{viscage_1}. \label{entropy}}
\end{figure}

This approach to the inclusion of bulk viscosity is highly testable in the sense that future measurements could rule it out. As mentioned above, the lifetime cannot be much less than 14 Gyr. Further, if the universe is flat, or nearly so, as seems likely, then we require $\Omega_M$ to be rather small. A close examination of Fig.\,\ref{viscage_1} near $\Omega_k=0$ reveals that the present supernova data rule out $\Omega_M=0.3$ at the 90\% level. On the other hand, improving the supernova contours is unlikely to rule out this model, even for a flat universe, as long as $\Omega_M\leq 0.1$ is allowed.

\acknowledgements We would like to thank Marcello Lissia, Vic Teplitz and Rocky Kolb for helpful comments. This research was supported in part by the National Science Foundation under Grant PHY-0070443 and by the United States Department of Energy under Contract No. DE-FG03-93ER40757.

\newpage
\begin{table}[h]
\centering
\begin{tabular}{p{1.0in}p{1.0in}p{1.0in}p{1.0in}p{1.0in}}
\toprule
\multicolumn{5}{c}{Supernova Cosmology Project}\\
& General & Flat & General & Flat \\
\colrule
$\Omega_M$         & 0.79  & 0.28$^{+0.09}_{-0.08}$ & 0.73 & 0.28$^{+0.09}_{-0.08}$ \\
$\Omega_{\Lambda}$ & 1.41  & $1-\Omega_M$           & 1.32 & $1-\Omega_M$           \\
$\mathcal{M}$      & 23.9  & 23.9                   &      &                        \\
$\chi^2$           & 56.9  & 57.7                   & 56.0 &                        \\[2pt]
\botrule
\end{tabular}
\caption{The first pair of entries corresponds to the fit obtained using Eq.\,(\ref{chisq}) and the second pair are the results from Ref.\cite{super}.
\label{fits}}
\end{table}

\begin{table}[h]
\centering
\begin{tabular}{p{1.0in}p{1.0in}p{1.0in}}
\toprule
\multicolumn{3}{c}{$w=-2/3$} \\
& General & Flat \\
\colrule
$\Omega_M$         & 0.54  & 0.08$^{+0.10}_{-0.08}$ \\
$\Omega_{\Lambda}$ & 2.33  & $1-\Omega_M$           \\
$\mathcal{M}$      & 23.9  & 24.0                   \\
$\chi^2$           & 56.81  & 58.2                   \\
[2pt]
\colrule
\multicolumn{3}{c}{$w=-0.73$}\\
& General & Flat  \\
\colrule
$\Omega_M$         & 0.63  & 0.13$^{+0.11}_{-0.10}$ \\
$\Omega_{\Lambda}$ & 2.10  & $1-\Omega_M$           \\
$\mathcal{M}$      & 23.9  & 24.0                   \\
$\chi^2$           & 56.80  & 58.1                   \\
[2pt]
\botrule
\end{tabular}
\caption{Fits to the data of Ref.\,\cite{super} for two constant
values of $w$ obtained using Eq.\,(\ref{chisq}) are shown. \label{quin1}}
\end{table}

\begin{table}[h]
\centering
\begin{tabular}{p{1.0in}p{1.0in}p{1.0in}}
\toprule
\multicolumn{3}{c}{$w=-2y/3$} \\
& General & Flat \\
\colrule
$\Omega_M$         & 0.18  & 0.0$^{+0.13}_{-0.18}$ \\
$\Omega_{\Lambda}$ & 2.33  & $1-\Omega_M$           \\
$\mathcal{M}$      & 23.9  & 24.0                   \\
$\chi^2$           & 56.9  & 58.0                   \\
[2pt]
\colrule
\multicolumn{3}{c}{$w=-y$}\\
& General & Flat  \\
\colrule
$\Omega_M$         & 0.55  & 0.21$\pm 0.08$ \\
$\Omega_{\Lambda}$ & 1.39  & $1-\Omega_M$           \\
$\mathcal{M}$      & 23.9  & 23.9                   \\
$\chi^2$           & 56.9  & 57.6                   \\
[2pt]
\botrule
\end{tabular}
\caption{Fits to the data of Ref.\,\cite{super} for the cases $w=-2y/3$ and
$w=-y$ obtained using Eq.\,(\ref{chisq}) are shown. \label{quin2}}
\end{table}

\begin{table}[h]
\centering
\begin{tabular}{p{1.0in}p{1.0in}p{1.0in}}
\toprule
\multicolumn{3}{c}{Bulk Viscosity} \\
\colrule
& General & Flat \\
\colrule
$\Omega_k$         & -1.36  & 0.0     \\
$A$                &  1.47  & 0.66    \\
$\mathcal{M}$      & 23.9   & 24.0    \\
$\chi^2$           & 56.9   & 58.0    \\
[2pt]
\botrule
\end{tabular}
\caption{Fits to the data of Ref.\,\cite{super} for the case of a dark
component consisting of an ideal fluid with a bulk viscosity $\zeta$ are shown.
\label{viscfit}}
\end{table}

\newpage

\end{document}